\documentclass[12pt,a4paper]{article}
\makeatletter
\def\eqnarray{%
\stepcounter{equation}%
\let\@currentlabel=\theequation
\global\@eqnswtrue
\global\@eqcnt\z@
\tabskip\@centering
\let\\=\@eqncr
$$\halign to \displaywidth\bgroup\@eqnsel\hskip\@centering
$\displaystyle\tabskip\z@{##}$&\global\@eqcnt\@ne
\hfil$\displaystyle{{}##{}}$\hfil
&\global\@eqcnt\tw@$\displaystyle\tabskip\z@{##}$\hfil
\tabskip\@centering&\llap{##}\tabskip\z@\cr}
\makeatother
\newcommand{\ket}[1]{{\vert{#1}\rangle}}
\newcommand{\bra}[1]{{\langle{#1}\vert}}

\begin{document}

\title{\sl N Level System with RWA and Analytical Solutions Revisited}
\author{
  Kazuyuki FUJII 
  \thanks{E-mail address : fujii@yokohama-cu.ac.jp },
  Kyoko HIGASHIDA 
  \thanks{E-mail address : s035577d@yokohama-cu.ac.jp },
  Ryosuke KATO 
  \thanks{E-mail address : s035559g@yokohama-cu.ac.jp }, 
  Yukako WADA 
  \thanks{E-mail address : s035588a@yokohama-cu.ac.jp }\\
  Department of Mathematical Sciences\\
  Yokohama City University\\
  Yokohama, 236--0027\\
  Japan
  }
\date{}
\maketitle
%
%
%
%
\begin{abstract}
  In this paper we consider a model of an atom with n energy levels 
  interacting with n(n-1)/2 external (laser) fields which is a natural 
  extension of two level system, and assume the rotating wave approximation 
  (RWA) from the beginning. 
  We revisit some construction of analytical solutions (which correspond to 
  Rabi oscillations) of the model in the general case 
  and examine it in detail in the case of three level system. 
\end{abstract}
%


%
%
%
%

\newpage

\section{Introduction}

Our purpose of this paper is to apply a method developed in this one 
to Quantum Computation. Quantum Computation (or Computer) is a challenging 
task in this century for not only physicists but also mathematicians. 
See for example \cite{LPS} as a general introduction. 

Quantum Computation is in a usual understanding based on qubits which are 
based on two level system (two energy levels or fundamental spins of atoms), 
See \cite{AE}, \cite{MSIII}, \cite{BR}, \cite{C-HDG} as for general theory of 
two level system. 
In a realistic image of Quantum Computer we need at least one hundred atoms. 
However then we meet a very severe problem called Decoherence which will 
destroy a superposition of quantum states in the process of unitary 
evolution of our system. See for example \cite{WHZ} or recent \cite{MFr3} 
as an introduction. At the present it is not easy to control Decoherence. 

By the way, an atom has in general infinitely many energy levels, while 
in a qubit method we use only two energy levels. 
We should use this possibility to reduce a number of atoms. 
We use $n$ energy levels from the ground state (it is not realistic to take 
all energy levels into consideration at the same time). We call this 
$n$ level system a qudit theory, see for example \cite{KF3}, \cite{KF6}, 
\cite{KuF}. 

In a qubit case we need Rabi oscillations to construct quantum logic gates. 
See \cite{KF1} as a simple introduction to quantum logic gates. 
Similarly we also need Rabi oscillations to construct them in a qudit space. 
However a general theory of Rabi oscillations in a $n$ level system has not 
been developed enough as far as we know. For three level system see 
\cite{MSIII}, \cite{BR} and \cite{MFr5}. 
Therefore we attempt such a theory in this paper. 

In this model we assume the rotating wave approximation (RWA) 
from the beginning, otherwise we cannot solve the model. However there is 
no problem on the approximation in the weak coupling regime (the coupling 
constant is small enough compared to the energy differences of an atom). 
We note that there is a problem to use this approximation in the strong 
coupling regime. 

Before stating our result let us make a review of Rabi oscillations 
(or coherent oscillations) in a two level system. 
For a quantum version of the Rabi oscillation see for example, \cite{C-HDG}, 
\cite{MFr2}, \cite{KF2}. 

Let $\{\sigma_{1}, \sigma_{2}, \sigma_{3}\}$ be Pauli matrices : 
\begin{equation}
\sigma_{1} = 
\left(
  \begin{array}{cc}
    0& 1 \\
    1& 0
  \end{array}
\right), \quad 
\sigma_{2} = 
\left(
  \begin{array}{cc}
    0& -i \\
    i& 0
  \end{array}
\right), \quad 
\sigma_{3} = 
\left(
  \begin{array}{cc}
    1& 0 \\
    0& -1
  \end{array}
\right), 
\end{equation}
and we set 
\[
\sigma_{+}=\frac{1}{2}(\sigma_{1}+i\sigma_{2})=
\left(
  \begin{array}{cc}
    0& 1 \\
    0& 0
  \end{array}
\right), \quad 
\sigma_{-}=\frac{1}{2}(\sigma_{1}-i\sigma_{2})=
\left(
  \begin{array}{cc}
    0& 0 \\
    1& 0
  \end{array}
\right). 
\]

Let us consider an atom with $2$ energy levels $E_{0}$ and $E_{1}$ (
$E_{1} > E_{0}$). 
Its Hamiltonian is in the diagonal form given as 
\begin{equation}
H_{0}=
\left(
  \begin{array}{cc}
    E_{0}& 0 \\
    0& E_{1}
  \end{array}
\right).
\end{equation}
This is rewritten as 
\[
H_{0}=
E_{0}
\left(
  \begin{array}{cc}
    1& 0 \\
    0& 1
  \end{array}
\right)+
(E_{1}-E_{0})
\left(
  \begin{array}{cc}
    0& 0   \\
    0& 1
  \end{array}
\right)
=
E_{0}{\bf 1}_{2}+\frac{\Delta}{2}\left({\bf 1}_{2}-\sigma_{3}\right),
\]
where $\Delta=E_{1}-E_{0}$ is an energy difference. 
Since we usually take no interest in constant terms, we can set 
\begin{equation}
\label{eq:2-energy-hamiltonian}
H_{0}=\frac{\Delta}{2}\left({\bf 1}_{2}-\sigma_{3}\right). 
\end{equation}

We consider an atom with two energy levels which interacts with external 
(periodic) field with $g\mbox{cos}(\omega t+\phi)$. 
In the following we set $\hbar=1$ for simplicity. 
The Hamiltonian in the dipole approximation is given by 
\begin{equation}
\label{eq:2-full-hamiltonian}
H=H_{0}+g \mbox{cos}(\omega t+\phi)\sigma_{1}
=\frac{\Delta}{2}\left({\bf 1}_{2}-\sigma_{3}\right)+
g\mbox{cos}(\omega t+\phi)\sigma_{1}, 
\end{equation}
where $\omega$ is the frequency of the external field, $g$ the coupling 
constant between the external field and the atom. This model is complicated 
enough, see \cite{MFr1}, \cite{MFr4}, \cite{KF5}, \cite{BaWr}, \cite{SGD}. 

In the following we set $\phi=0$ for simplicity and assume the rotating wave 
approximation (which neglects the fast oscillating terms), namely 
\[
\mbox{cos}(\omega t)
=\frac{1}{2}(\mbox{e}^{i\omega t}+\mbox{e}^{-i\omega t})
=\frac{1}{2}\mbox{e}^{i\omega t}(1+\mbox{e}^{-2i\omega t})
\approx \frac{1}{2}\mbox{e}^{i\omega t},
\]
and 
\[
\mbox{cos}(\omega t)\sigma_{1}
=
\left(
  \begin{array}{cc}
                     0  & \mbox{cos}(\omega t) \\
    \mbox{cos}(\omega t)& 0
  \end{array}
\right)
\approx 
\frac{1}{2}
\left(
  \begin{array}{cc}
                       0 & \mbox{e}^{i\omega t} \\
    \mbox{e}^{-i\omega t}& 0
  \end{array}
\right), 
\]
therefore the Hamiltonian is given by 
\begin{equation}
\label{eq:sub-hamiltonian}
H=
\frac{\Delta}{2}\left({\bf 1}_{2}-\sigma_{3}\right)+\frac{g}{2}
\left(\mbox{e}^{i\omega t}\sigma_{+}+\mbox{e}^{-i\omega t}\sigma_{-}\right)
\equiv 
\frac{\Delta}{2}\left({\bf 1}_{2}-\sigma_{3}\right)+
g\left(\mbox{e}^{i\omega t}\sigma_{+}+\mbox{e}^{-i\omega t}\sigma_{-}\right)
\end{equation}
by the redefinition of $g$ ($g/2 \longrightarrow g$). 
It is explicitly 
\begin{equation}
\label{eq:reduction-matrix}
H=
\left(
  \begin{array}{cc}
    0& g\mbox{e}^{i\omega t}         \\
    g\mbox{e}^{-i\omega t}& \Delta
  \end{array}
\right). 
\end{equation}

We would like to solve the Schr{\" o}dinger equation 
\begin{equation}
\label{eq:schrodinger-equation}
i\frac{d}{dt}\Psi=H\Psi.
\end{equation}
For that let us decompose $H$ in (\ref{eq:reduction-matrix}) into 
\begin{equation}
\label{eq:decomposition}
\left(
  \begin{array}{cc}
    0& g\mbox{e}^{i\omega t}         \\
    g\mbox{e}^{-i\omega t}& \Delta
  \end{array}
\right)
=
\left(
  \begin{array}{cc}
    1&         \\
     & \mbox{e}^{-i\omega t}
  \end{array}
\right)
\left(
  \begin{array}{cc}
    0& g         \\
    g& \Delta
  \end{array}
\right)
\left(
  \begin{array}{cc}
    1&         \\
     & \mbox{e}^{i\omega t}
  \end{array}
\right),
\end{equation}
so if we set 
\begin{equation}
\label{eq:transformation}
\Phi=
\left(
  \begin{array}{cc}
    1&         \\
     & \mbox{e}^{i\omega t}
  \end{array}
\right)
\Psi
\quad \Longleftrightarrow \quad 
\Psi=
\left(
  \begin{array}{cc}
    1&         \\
     & \mbox{e}^{-i\omega t}
  \end{array}
\right)
\Phi
\end{equation}
then it is not difficult to see 
\begin{equation}
\label{eq:schrodinger-equation-2}
i\frac{d}{dt}\Phi=
\left(
  \begin{array}{cc}
    0& g              \\
    g& \Delta-\omega 
  \end{array}
\right)
\Phi,
\end{equation}
which is easily solved. For simplicity we set the resonance condition 
\begin{equation}
\label{eq:resonance}
\Delta=\omega,
\end{equation}
then the solution of (\ref{eq:schrodinger-equation-2}) is 
\[
\Phi(t)=
\mbox{exp}
\left\{-igt
\left(
  \begin{array}{cc}
    0& 1   \\
    1& 0 
  \end{array}
\right)
\right\}
\Phi(0)=
\left(
  \begin{array}{cc}
    \mbox{cos}(gt)& -i\mbox{sin}(gt)   \\
    -i\mbox{sin}(gt)& \mbox{cos}(gt) 
  \end{array}
\right)
\Phi(0).
\]
As a result, the solution of the equation (\ref{eq:schrodinger-equation}) is 
given as 
\begin{equation}
\Psi(t)
=
\left(
  \begin{array}{cc}
    1&         \\
     & \mbox{e}^{-i\omega t}
  \end{array}
\right)
\Phi(t)
=
\left(
  \begin{array}{cc}
    1&         \\
     & \mbox{e}^{-i\omega t}
  \end{array}
\right)
\left(
  \begin{array}{cc}
    \mbox{cos}(gt)& -i\mbox{sin}(gt)   \\
    -i\mbox{sin}(gt)& \mbox{cos}(gt) 
  \end{array}
\right)
\Phi(0)
\end{equation}
by (\ref{eq:transformation}). If we choose 
$
\Phi(0)=\left(1,\ 0\right)^{T}
$
 as an initial condition, then 
\begin{equation}
\Psi(t)=
\left(
  \begin{array}{c}
    \mbox{cos}(gt)                         \\
  -i\mbox{e}^{-i\omega t}\mbox{sin}(gt)
  \end{array}
\right).
\end{equation}
This is a well--known model of the Rabi oscillation (or coherent 
oscillation).

\section{N Level System $\cdots$ General Theory}

In general, an atom has an infinitely many energy levels, however it is not 
realistic to consider all of them at the same time. Therefore we take 
only $n$ energy levels from the ground state into consideration 
($E_{n-1} > \cdots > E_{1} > E_{0}$). Then from the lesson in the preceding 
section the energy Hamiltonian can be 
written as 
\begin{equation}
\label{eq:}
H_{0}=
\left(
  \begin{array}{cccccc}
    0&           &           &       &      &              \\
     &\Delta_{1} &           &       &      &              \\
     &           &\Delta_{2} &       &      &              \\
     &           &           & \cdot &      &              \\
     &           &           &       &\cdot &              \\
     &           &           &       &      & \Delta_{n-1} 
  \end{array}
\right),
\end{equation}
where $\Delta_{j}=E_{j}-E_{0}$ for $1 \leq j \leq  n-1$. For this atom 
we consider the interaction with $n(n-1)/2$ independent external (laser) 
fields corresponding to every energy difference ($E_{j}-E_{i}$ for $j > i$). 
Then the interaction term assuming the RWA from the beginning 
is given as 
\begin{equation}
\label{eq:}
V=
\left(
  \begin{array}{ccccccc}
    0& \mbox{e}^{i\omega_{01}t} &\mbox{e}^{i\omega_{02}t} &\cdot &\qquad \cdot 
    &\cdot & \mbox{e}^{i\omega_{0,n-1}t}\\
   \mbox{e}^{-i\omega_{01}t}&0 & \mbox{e}^{i\omega_{12}t} &\cdot &\qquad \cdot 
   &\cdot & \mbox{e}^{i\omega_{1,n-1}t} \\
   \mbox{e}^{-i\omega_{02}t}& \mbox{e}^{-i\omega_{12}t} &0 & \cdot
   &\qquad \cdot &\cdot & \mbox{e}^{i\omega_{2,n-1}t}  \\
   \cdot& \cdot & \cdot & \cdot &\qquad \cdot& \cdot  &\cdot \\
   \cdot& \cdot  & \cdot &\cdot &\qquad \cdot & \cdot & \cdot    \\
   \cdot& \cdot  & \cdot &\cdot &\qquad \cdot & 0 & 
   \mbox{e}^{i\omega_{n-2,n-1}t} \\
   \mbox{e}^{-i\omega_{0,n-1}t}& \mbox{e}^{-i\omega_{1,n-1}t} & 
   \mbox{e}^{-i\omega_{2,n-1}t} & \cdot &\qquad \cdot & 
   \mbox{e}^{-i\omega_{n-2,n-1}t} & 0 
  \end{array}
\right),
\end{equation}
so the full Hamiltonian is 
\begin{equation}
\label{eq:}
H=H_{0}+gV, 
\end{equation}
where $g$ is the coupling constant. Here we set 
\[
\omega_{1}=\omega_{01},\ \omega_{2}=\omega_{12},\ \cdots, \ 
\omega_{n-2}=\omega_{n-3,n-2},\ \omega_{n-1}=\omega_{n-2,n-1} 
\]
for simplicity. 
We would like to solve the Schr{\" o}dinger equation 
\begin{equation}
\label{eq:schrodinger-equation-general}
i\frac{d}{dt}\Psi=H\Psi.
\end{equation}
Similarly in (\ref{eq:decomposition}) let us decompose $H$ : 
For 
\begin{equation}
\label{eq:U}
U=
\left(
  \begin{array}{ccccccc}
    1&                             & & & & &  \\
     & \mbox{e}^{i\omega_{1}t}     & & & &  & \\
     &   & \mbox{e}^{i(\omega_{1}+\omega_{2})t} & & & &  \\
     & & & \cdot  & & & \\
     & & & & \qquad \cdot &  & \\
     & & & & & \mbox{e}^{i(\omega_{1}+\omega_{2}+\cdots +\omega_{n-2})t} & \\
     & & & & & & \mbox{e}^{i(\omega_{1}+\omega_{2}+\cdots +\omega_{n-1})t}
  \end{array}
\right)
\end{equation}
it is easy to see 
\begin{equation}
\label{eq:}
H_{U}\equiv U^{\dagger}HU=
\left(
  \begin{array}{ccccccc}
    0& g & g\mbox{e}^{i\epsilon_{02}t} &\cdot &\quad \cdot 
    & g\mbox{e}^{i\epsilon_{0,n-2}t} & g\mbox{e}^{i\epsilon_{0,n-1}t}\\
    g& \Delta_{1} & g &\cdot &\quad \cdot 
    & g\mbox{e}^{i\epsilon_{1,n-2}t} & g\mbox{e}^{i\epsilon_{1,n-1}t} \\
    g\mbox{e}^{-i\epsilon_{02}t}& g &\Delta_{2} & \cdot
    &\quad \cdot & g\mbox{e}^{i\epsilon_{2,n-2}t} & 
    g\mbox{e}^{i\epsilon_{2,n-1}t}  \\
    \cdot& \cdot & \cdot & \cdot &\quad \cdot& \cdot  &\cdot \\
    \cdot& \cdot  & \cdot &\cdot &\quad \cdot & \cdot & \cdot    \\
    g\mbox{e}^{-i\epsilon_{0,n-2}t} & g\mbox{e}^{-i\epsilon_{1,n-2}t} & 
    g\mbox{e}^{-i\epsilon_{2,n-2}t} & \cdot &\quad \cdot & \Delta_{n-2} & g \\
    g\mbox{e}^{-i\epsilon_{0,n-1}t}& g\mbox{e}^{-i\epsilon_{1,n-1}t} & 
    g\mbox{e}^{-i\epsilon_{2,n-1}t} & \cdot &\quad \cdot & g & \Delta_{n-1}
  \end{array}
\right),
\end{equation}
where 
\[
\epsilon_{ij}=\omega_{ij}-(\omega_{i+1}+\omega_{i+2}+\cdots+\omega_{j})
\]
for $j-i\geq 2$. 

By setting 
\[
\tilde{\Psi}=U\Psi\ \Longleftrightarrow \ \Psi=U^{\dagger}\tilde{\Psi}
\]
it is not difficult to see 
\begin{equation}
\label{eq:second-schrodinger-equation}
i\frac{d}{dt}\tilde{\Psi}=\tilde{H}_{U}\tilde{\Psi},
\end{equation}
where 
\begin{eqnarray}
\label{eq:}
&&\tilde{H}_{U}=          \nonumber \\
&&\left(
  \begin{array}{ccccccc}
    0& g & g\mbox{e}^{i\epsilon_{02}t} &\cdot & \cdot 
    & g\mbox{e}^{i\epsilon_{0,n-2}t} & g\mbox{e}^{i\epsilon_{0,n-1}t}  \\
    g& \Delta_{1}-\omega_{1} & g &\cdot & \cdot 
    & g\mbox{e}^{i\epsilon_{1,n-2}t} & g\mbox{e}^{i\epsilon_{1,n-1}t} \\
    g\mbox{e}^{-i\epsilon_{02}t}& g &\Delta_{2}-(\omega_{1}+\omega_{2})& \cdot
    & \cdot & g\mbox{e}^{i\epsilon_{2,n-2}t} & g\mbox{e}^{i\epsilon_{2,n-1}t}\\
    \cdot& \cdot & \cdot & \cdot & \cdot& \cdot  &\cdot \\
    \cdot& \cdot  & \cdot &\cdot & \cdot & \cdot & \cdot    \\
    g\mbox{e}^{-i\epsilon_{0,n-2}t}& g\mbox{e}^{-i\epsilon_{1,n-2}t} & 
    g\mbox{e}^{-i\epsilon_{2,n-2}t} & \cdot & \cdot &
    \Delta_{n-2}-\sum_{l=1}^{n-2}\omega_{l} & g \\
    g\mbox{e}^{-i\epsilon_{0,n-1}t}& g\mbox{e}^{-i\epsilon_{1,n-1}t} & 
    g\mbox{e}^{-i\epsilon_{2,n-1}t} & \cdot & \cdot & g & 
    \Delta_{n-1}-\sum_{l=1}^{n-1}\omega_{l}
  \end{array}
\right).                  \nonumber \\
&&{} 
\end{eqnarray}
At this stage we set the resonance conditions 
\begin{eqnarray}
&&\Delta_{1}=\omega_{1},\ \Delta_{2}=\omega_{1}+\omega_{2}, \ \cdots,\ 
\Delta_{n-2}=\sum_{l=1}^{n-2}\omega_{l},\ 
\Delta_{n-1}=\sum_{l=1}^{n-1}\omega_{l}      \nonumber \\
\Longleftrightarrow \ &&\omega_{j}=E_{j}-E_{j-1}\ (j=1,\ 2,\ \cdots,\ n-1) 
\end{eqnarray}
to make the situation simpler. Therefore $\tilde{H}_{U}$ can be written as 
\[
\tilde{H}_{U}=gC+gR, 
\]
where 
\begin{equation}
\label{eq:semi-diagonal}
C=
\left(
  \begin{array}{ccccccc}
    0 & 1 &   &   &   &   &              \\
    1 & 0 & 1 &   &   &   &              \\
      & 1 & 0 & 1 &       &              \\
      &   & \cdot & \cdot & \cdot  &   & \\
      &   &  & \cdot & \cdot  & \cdot  & \\
      &   &   &   & 1 & 0 & 1            \\
      &   &   &   &   & 1 & 0
  \end{array}
\right)             
\end{equation}
and 
\begin{equation}
\label{eq:non-diagonal}
R(t)=
  \left(
  \begin{array}{ccccccc}
    0& 0 & \mbox{e}^{i\epsilon_{02}t} &\cdot & \cdot 
    &\mbox{e}^{i\epsilon_{0,n-2}t} & \mbox{e}^{i\epsilon_{0,n-1}t} \\ 
    0& 0 & 0 &\cdot & \cdot & \mbox{e}^{i\epsilon_{1,n-2}t} & 
    \mbox{e}^{i\epsilon_{1,n-1}t}  \\
    \mbox{e}^{-i\epsilon_{02}t}& 0 & 0 & \cdot & \cdot & 
    \mbox{e}^{i\epsilon_{2,n-2}t} & \mbox{e}^{i\epsilon_{2,n-1}t}  \\
    \cdot& \cdot & \cdot & \cdot & \cdot& \cdot  &\cdot     \\
    \cdot& \cdot  & \cdot &\cdot & \cdot & \cdot & \cdot    \\
    \mbox{e}^{-i\epsilon_{0,n-2}t}& \mbox{e}^{-i\epsilon_{1,n-2}t} & 
    \mbox{e}^{-i\epsilon_{2,n-2}t} & \cdot & \cdot & 0 & 0    \\
    \mbox{e}^{-i\epsilon_{0,n-1}t}& \mbox{e}^{-i\epsilon_{1,n-1}t} & 
    \mbox{e}^{-i\epsilon_{2,n-1}t} & \cdot & \cdot & 0 & 0 
  \end{array}
\right).       
\end{equation}
Therefore the Schr{\" o}dinger equation becomes 
\begin{equation}
\label{eq:schrodinger-perturbation}
i\frac{d}{dt}\tilde{\Psi}=(gC+gR)\tilde{\Psi}.
\end{equation}

Now it is easy to solve 
\[
i\frac{d}{dt}\tilde{\Psi}=gC\tilde{\Psi},
\]
which solution is 
\[
\tilde{\Psi}(t)=\mbox{exp}(-igtC)\tilde{\Psi}(0).
\]
By making use of the method of constant variation 
\[
\tilde{\Psi}(t)=\mbox{exp}(-igtC)\tilde{\Psi}(0)\longrightarrow 
\tilde{\Psi}(t)=\mbox{exp}(-igtC)\phi(t)
\]
and substituting this into (\ref{eq:schrodinger-perturbation}) we have the 
equation
\begin{eqnarray}
\label{eq:interaction-picture}
i\frac{d}{dt}\phi&=&g\mbox{exp}(igtC)R(t)\mbox{exp}(-igtC)\phi
\equiv gA(t)\phi, \quad \mbox{where}  \\
\label{eq:A(t)}
A(t)&=&\mbox{exp}(igtC)R(t)\mbox{exp}(-igtC).
\end{eqnarray}
For this equation a formal solution called the Dyson series\footnote{In 
Mathematics it was called the Peano series a long time ago, \cite{AA}} 
has been known 
\begin{eqnarray}
\label{eq:dyson-series}
\phi(t)
&=&:\mbox{exp}\left(-ig\int_{0}^{t}A(s)ds \right):\phi(0)  \nonumber \\
&\equiv&
\left[
{\bf 1}-ig\int_{0}^{t}A(s)ds-
g^{2}\int_{0}^{t}\left\{A(s)\int_{0}^{s}A(u)du\right\}ds +\cdots 
\right]\phi(0).
\end{eqnarray}
However it is usually almost impossible to perform integrals. 
Anyway, our formal solution of the equation 
(\ref{eq:second-schrodinger-equation}) becomes 
\begin{equation}
\label{eq:formal-solution}
\tilde{\Psi}(t)=
\mbox{exp}(-igtC):\mbox{exp}\left(-ig\int_{0}^{t}A(s)ds \right):{\Psi} 
\end{equation}
where ${\Psi}$ is a constant vector. As a conclusion, a formal solution 
of (\ref{eq:schrodinger-equation-general}) is given by 
\begin{equation}
\label{eq:full-formal-solution}
\Psi(t)=U(t)^{\dagger}\tilde{\Psi}(t)=U(t)^{\dagger}\mbox{exp}(-igtC)
:\mbox{exp}\left(-ig\int_{0}^{t}A(s)ds \right):\Psi
\end{equation}
with $U(t)$ in (\ref{eq:U}). 

Lastly let us calculate $\mbox{exp}(-igtC)$ in 
(\ref{eq:full-formal-solution}). For that purpose we need to make $C$ in 
(\ref{eq:semi-diagonal}) a diagonal form. 
First let us calculate eigenvalues. The characteristic equation of $C$ is 
\begin{equation}
f_{n}(\lambda)\equiv 
\left|
  \begin{array}{ccccccc}
    \lambda & -1 &   &   &   &   &              \\
    -1& \lambda & -1 &   &   &   &              \\
      & -1 & \lambda & -1 &       &              \\
      &   & \cdot & \cdot & \cdot  &   &        \\
      &   &  & \cdot & \cdot  & \cdot  &        \\
      &   &   &   & -1 & \lambda & -1            \\
      &   &   &   &    & -1      & \lambda
  \end{array}
\right|=0.             
\end{equation}
By the Laplace expansion with respect to the first column we have a recurrent 
equation 
\begin{equation}
f_{n}(\lambda)=\lambda f_{n-1}(\lambda)-f_{n-2}(\lambda)
\quad \mbox{and}\quad 
f_{2}(\lambda)=\lambda^{2}-1,\ f_{3}(\lambda)=\lambda^{3}-2\lambda.
\end{equation}
For example, 
$
f_{4}(\lambda)=\lambda^{4}-3\lambda^{2}+1
$
. From these equations it is not easy to conjecture the eigenvalues. 
However we know a good method ! We set 
\begin{equation}
\lambda=x+\frac{1}{x}=x+x^{-1}.
\end{equation}
Then for example, 
\[
f_{2}(x)=x^{2}+1+x^{-2},\ 
f_{3}(x)=x^{3}+x^{1}+x^{-1}+x^{-3},\ 
f_{4}(x)=x^{4}+x^{2}+1+x^{-2}+x^{-4}.
\]
In general 
\begin{equation}
f_{n}(x)=\sum_{j=0}^{n}x^{-n+2j}=x^{-n}\frac{x^{2n+2}-1}{x^{2}-1}.
\end{equation}
The solutions of $f_{n}(x)=0$ are $\{\mbox{exp}(\sqrt{-1}\pi j/(n+1))\ |\ 
j=1,2,\cdots, n, n+2,\cdots, 2n+1\}$ ($\sqrt{-1}=i$) because $x\ne \pm 1$. 
Therefore our solutions are 
\begin{equation}
\label{eq:eigenvalues}
\lambda_{j}=x_{j}+x_{j}^{-1}=2\mbox{cos}\left(\frac{\pi j}{n+1}\right) 
\quad \mbox{for}\quad j=1,2,\cdots,n 
\end{equation}
by removing the overlap. 

Next let us calculate an eigenfunction corresponding to $\lambda_{j}$ 
\[
C\ket{j}
=\lambda_{j}\ket{j}
=2\mbox{cos}\left(\frac{\pi j}{n+1}\right)\ket{j}
\]
for $j=1, 2, \cdots , n$. If we define 
\begin{equation}
\ket{j}=\sqrt{\frac{2}{n+1}}
  \left(
  \mbox{sin}\left(\frac{\pi j}{n+1}\right), 
  \mbox{sin}\left(\frac{\pi 2j}{n+1}\right), \cdots, 
  \mbox{sin}\left(\frac{\pi kj}{n+1}\right), \cdots, 
  \mbox{sin}\left(\frac{\pi nj}{n+1}\right)
  \right)^{T},
\end{equation}
then it is not difficult to see 
\begin{equation}
C\ket{j}=2\mbox{cos}\left(\frac{\pi j}{n+1}\right)\ket{j}\quad 
\mbox{and}\quad \langle{j|k}\rangle=\delta_{jk}.
\end{equation}
We leave the proof to the readers. Now we define 
\begin{equation}
O=(O_{jk})\quad \mbox{where}\quad 
O_{jk}=\sqrt{\frac{2}{n+1}}\mbox{sin}\left(\frac{\pi jk}{n+1}\right)
\quad \Longrightarrow \quad OO^{T}={\bf 1}_{n}
\end{equation}
and 
\begin{equation}
D=(D_{jk})\quad \mbox{where}\quad 
D_{jk}=2\mbox{cos}\left(\frac{\pi j}{n+1}\right)\delta_{jk},
\end{equation}
then we can diagonalize $C$ like 
\begin{equation}
C=ODO^{T}.
\end{equation}
Therefore we finally obtain 
\begin{equation}
\label{eq:n=n}
\mbox{exp}(-igtC)=O\mbox{exp}(-igtD)O^{T}
\end{equation}
or in components 
\begin{equation}
\label{eq:n=n components}
\mbox{exp}(-igtC)_{jk}=\frac{2}{n+1}\sum_{l=1}^{n}
\mbox{exp}\left(-2igt\mbox{cos}\left(\frac{\pi l}{n+1}\right)\right)
\mbox{sin}\left(\frac{\pi jl}{n+1}\right)
\mbox{sin}\left(\frac{\pi kl}{n+1}\right).
\end{equation}

In the case of $n=3$ let us write down $\mbox{exp}(-igtC)$ explicitly 
for the sake of latter convenience. 
\begin{equation}
\label{eq:n=3}
\mbox{exp}(-igtC)=
\frac{1}{2}
\left(
  \begin{array}{ccc}
     1+\mbox{cos}(\sqrt{2}gt)& -i\sqrt{2}\mbox{sin}(\sqrt{2}gt)& 
    -1+\mbox{cos}(\sqrt{2}gt)  \\
     -i\sqrt{2}\mbox{sin}(\sqrt{2}gt)& 2\mbox{cos}(\sqrt{2}gt)& 
     -i\sqrt{2}\mbox{sin}(\sqrt{2}gt)  \\
     -1+\mbox{cos}(\sqrt{2}gt)& -i\sqrt{2}\mbox{sin}(\sqrt{2}gt)&
      1+\mbox{cos}(\sqrt{2}gt)
  \end{array}
\right). 
\end{equation}

\vspace{5mm}
For (\ref{eq:n=n}) with (\ref{eq:n=n components}) we would like to calculate 
the Dyson series (\ref{eq:dyson-series}). However it is almost impossible in 
the general case, so in the last section we calculate the next--leading term 
in the case of $n=3$ by making use of (\ref{eq:n=3}).

\section{N Level System $\cdots$ An Exact Solution}

In this section we present an exact solution in the general case under 
some ``consistency condition" (in our terminology). 

In section 2 we consider a very special case : namely, 
in (\ref{eq:non-diagonal}) we take 
\begin{equation}
\epsilon_{ij}=\omega_{ij}-(\omega_{i+1}+\omega_{i+2}+\cdots+\omega_{j})=0
\end{equation}
for all $j-i \geq 2$. We call this a consistency condition. 
Then $\tilde{H}_{U}$ becomes a constant matrix ! 
\begin{equation}
\tilde{H}_{U}=gC+gR=
g
\left(
  \begin{array}{ccccccc}
    0 & 1 & 1 & \cdot  & \cdot & 1  &  1                    \\
    1 & 0 & 1 &  \cdot & \cdot & 1  &  1                    \\
    1 & 1 & 0 & \cdot  & \cdot & 1  &  1                    \\
    \cdot & \cdot  & \cdot & \cdot & \cdot & \cdot & \cdot  \\
    \cdot & \cdot  & \cdot & \cdot & \cdot & \cdot & \cdot  \\
    1 & 1 & 1 & \cdot  & \cdot & 0  &  1                    \\
    1 & 1 & 1 & \cdot  & \cdot & 1  &  0
  \end{array}
\right)\equiv gQ,
\end{equation}
so it is easy to solve the equation (\ref{eq:schrodinger-perturbation}) 
\[
i\frac{d}{dt}\tilde{\Psi}=g(C+R)\tilde{\Psi}=gQ\tilde{\Psi}.
\]
The solution is formally 
\[
\tilde{\Psi}(t)=\mbox{exp}(-igtQ)\tilde{\Psi}(0). 
\]
Let us calculate $\mbox{exp}(-igtQ)$ explicitly. If we define 
\[
\ket{{\bf 1}}=(1,1,\cdots,1,1)^{T},
\]
then it is easy to see 
$
Q=\ket{{\bf 1}}\bra{{\bf 1}}-{\bf 1}_{n}
$, 
so 
\[
\tilde{\Psi}(t)=e^{igt}
\mbox{exp}(-igt\ket{{\bf 1}}\bra{{\bf 1}})\tilde{\Psi}(0). 
\]
Since $\langle{\bf 1}|{\bf 1}\rangle =n$, 
\[
\left(\ket{{\bf 1}}\bra{{\bf 1}}\right)^{k}
=n^{k-1}\ket{{\bf 1}}\bra{{\bf 1}},
\]
so that 
\begin{eqnarray}
\mbox{exp}(-igt\ket{{\bf 1}}\bra{{\bf 1}})
&=&{\bf 1}_{n}+
\sum_{k=1}^{\infty}
\frac{(-igt)^{k}}{k!}\left(\ket{{\bf 1}}\bra{{\bf 1}}\right)^{k}
={\bf 1}_{n}+
\sum_{k=1}^{\infty}
\frac{(-igt)^{k}}{k!}n^{k-1} \ket{{\bf 1}}\bra{{\bf 1}}  \nonumber \\
&=&
{\bf 1}_{n}+\frac{1}{n}
\left\{
\sum_{k=0}^{\infty}\frac{(-ingt)^{k}}{k!}-1
\right\}
\ket{{\bf 1}}\bra{{\bf 1}} 
={\bf 1}_{n}+\frac{\mbox{exp}(-ingt)-1}{n}\ket{{\bf 1}}\bra{{\bf 1}}.
\nonumber \\
&&{}
\end{eqnarray}
Therefore we obtain 
\begin{equation}
\tilde{\Psi}(t)=e^{igt}
\left\{
{\bf 1}_{n}+\frac{\mbox{exp}(-ingt)-1}{n}\ket{{\bf 1}}\bra{{\bf 1}}
\right\}
\tilde{\Psi}(0).
\end{equation}
Here if we choose $\tilde{\Psi}(0)=(1,0,\cdots,0,0)^{T}=\ket{0}$, then 
\[
\tilde{\Psi}(t)=e^{igt}
\left\{
\ket{0}+\frac{\mbox{exp}(-ingt)-1}{n}\ket{{\bf 1}}
\right\}.
\]
As a result, the solution that is looking for is just 
\begin{equation}
\Psi(t)=e^{igt}
U(t)^{\dagger}
\left\{
\ket{0}+\frac{\mbox{exp}(-ingt)-1}{n}\ket{{\bf 1}}
\right\} 
\end{equation}
with (\ref{eq:U}). In particular, in the case of $n=3$ the solution is 
\begin{eqnarray}
\Psi(t)&=&e^{igt}
\left(
  \begin{array}{ccc}
    1&                           &                 \\
     & \mbox{e}^{-i\omega_{1}t}  &                 \\
     &   & \mbox{e}^{-i(\omega_{1}+\omega_{2})t}   \\
  \end{array}
\right)
\left(
  \begin{array}{c}
    \frac{\mbox{exp}(-i3gt)+2}{3}  \\
    \frac{\mbox{exp}(-i3gt)-1}{3}  \\
    \frac{\mbox{exp}(-i3gt)-1}{3}
  \end{array}
\right)           \nonumber \\
&=&
\left(
  \begin{array}{c}
   \qquad \ \  e^{igt}\frac{\mbox{exp}(-i3gt)+2}{3}  \\
   \ \ \ e^{i(g-\omega_{1})t}\frac{\mbox{exp}(-i3gt)-1}{3}  \\
   e^{i(g-\omega_{1}-\omega_{2})t}\frac{\mbox{exp}(-i3gt)-1}{3}
  \end{array}
\right).
\end{eqnarray}

\section{Three Level System $\cdots$ An Approximate Solution}

We calculate the case of $n=3$ in more detail. For that purpose let us 
list $U(t)^{\dagger}$, $\mbox{exp}(-igtC)$ and $R(t)$ once more : 
\[
U(t)^{\dagger}=
\left(
  \begin{array}{ccc}
    1&                           &                 \\
     & \mbox{e}^{-i\omega_{1}t}  &                 \\
     &   & \mbox{e}^{-i(\omega_{1}+\omega_{2})t}   \\
  \end{array}
\right)
\]
and 
\[
\mbox{exp}(-igtC)=
\frac{1}{2}
\left(
  \begin{array}{ccc}
     1+\mbox{cos}(\sqrt{2}gt)& -i\sqrt{2}\mbox{sin}(\sqrt{2}gt)& 
    -1+\mbox{cos}(\sqrt{2}gt)  \\
     -i\sqrt{2}\mbox{sin}(\sqrt{2}gt)& 2\mbox{cos}(\sqrt{2}gt)& 
     -i\sqrt{2}\mbox{sin}(\sqrt{2}gt)  \\
     -1+\mbox{cos}(\sqrt{2}gt)& -i\sqrt{2}\mbox{sin}(\sqrt{2}gt)&
      1+\mbox{cos}(\sqrt{2}gt)
  \end{array}
\right) 
\]
and 
\[
R(t)=
\left(
  \begin{array}{ccc}
    0 & 0 & \mbox{exp}(i\epsilon t)  \\
    0 & 0 & 0                        \\
    \mbox{exp}(-i\epsilon t) & 0 & 0
  \end{array}
\right) 
\]
where $\epsilon\equiv \epsilon_{02}=\omega_{02}-(\omega_{1}+\omega_{2})$ 
for simplicity. 

First of all we must calculate $A(t)$ in (\ref{eq:A(t)}). 
This calculation is tedious even in the case of $n=3$. The result is 
\begin{equation}
A(t)=\mbox{exp}(igtC)R(t)\mbox{exp}(-igtC)=
\frac{1}{2}
\left(
  \begin{array}{ccc}
    a_{11}& a_{12} & a_{13}  \\
    a_{21}& a_{22} & a_{23}  \\
    a_{31}& a_{32} & a_{33}  
  \end{array}
\right) 
\end{equation}
where 
\begin{eqnarray}
a_{11}&=&-\mbox{sin}^{2}(\sqrt{2}gt)\mbox{cos}(\epsilon t), 
\nonumber \\ 
a_{12}&=&\sqrt{2}\mbox{sin}(\sqrt{2}gt)\mbox{sin}(\epsilon t)
-i\sqrt{2}\mbox{sin}(\sqrt{2}gt)\mbox{cos}(\sqrt{2}gt)\mbox{cos}(\epsilon t), 
\nonumber \\ 
a_{13}&=&\left(1+\mbox{cos}^{2}(\sqrt{2}gt)\right)\mbox{cos}(\epsilon t)
+i2\mbox{cos}(\sqrt{2}gt)\mbox{sin}(\epsilon t), 
\nonumber \\
a_{21}&=&\sqrt{2}\mbox{sin}(\sqrt{2}gt)\mbox{sin}(\epsilon t)
+i\sqrt{2}\mbox{sin}(\sqrt{2}gt)\mbox{cos}(\sqrt{2}gt)\mbox{cos}(\epsilon t), 
\nonumber \\ 
a_{22}&=&2\mbox{sin}^{2}(\sqrt{2}gt)\mbox{cos}(\epsilon t), 
\nonumber \\
a_{23}&=&-\sqrt{2}\mbox{sin}(\sqrt{2}gt)\mbox{sin}(\epsilon t)
+i\sqrt{2}\mbox{sin}(\sqrt{2}gt)\mbox{cos}(\sqrt{2}gt)\mbox{cos}(\epsilon t), 
\nonumber \\
a_{31}&=&\left(1+\mbox{cos}^{2}(\sqrt{2}gt)\right)\mbox{cos}(\epsilon t)
-i2\mbox{cos}(\sqrt{2}gt)\mbox{sin}(\epsilon t), 
\nonumber \\
a_{32}&=&-\sqrt{2}\mbox{sin}(\sqrt{2}gt)\mbox{sin}(\epsilon t)
-i\sqrt{2}\mbox{sin}(\sqrt{2}gt)\mbox{cos}(\sqrt{2}gt)\mbox{cos}(\epsilon t), 
\nonumber \\
a_{33}&=&-\mbox{sin}^{2}(\sqrt{2}gt)\mbox{cos}(\epsilon t).  
\nonumber 
\end{eqnarray}

Next we must calculate the Dyson series (\ref{eq:dyson-series}) with 
$A(t)$ above. To calculate all terms in (\ref{eq:dyson-series}) 
is of course impossible, so we calculate up to the next--leading 
term : 
\begin{equation}
\label{eq:approx}
\phi(t)\approx 
\left[
{\bf 1}-ig\int_{0}^{t}A(s)ds
\right]\phi(0)
\end{equation}
with $\phi(0)$ being the ground state $\phi(0)=(1,0,0)^{T}$. 
By the way, it is convenient for us to calculate 
\begin{equation}
\label{eq:approx-2}
\mbox{exp}(-igtC)\phi(t)\approx 
\left[\mbox{exp}(-igtC)
-ig\int_{0}^{t}\mbox{exp}(ig(s-t)C)R(s)\mbox{exp}(-igsC)ds
\right]\phi(0)
\end{equation}
rather than (\ref{eq:approx}) itself. The result is 
\begin{equation}
\mbox{exp}(-igtC)\phi(t)=
\left(x_{1}(t), x_{2}(t), x_{3}(t)\right)^{T}, 
\end{equation}
where 
\begin{eqnarray}
x_{1}(t)&=&\frac{1+\mbox{cos}(\sqrt{2}gt)}{2}-
\frac{ig}{4}\left(-2+\mbox{cos}(\sqrt{2}gt)\right)
\frac{\mbox{sin}(\epsilon t)}{\epsilon}-       \nonumber \\
&{}&\frac{ig}{8}
\left\{
\frac{\mbox{sin}(\sqrt{2}gt)+\mbox{sin}((\sqrt{2}g+\epsilon)t)}
{2\sqrt{2}g+\epsilon}+
\frac{\mbox{sin}(\sqrt{2}gt)+\mbox{sin}((\sqrt{2}g-\epsilon)t)}
{2\sqrt{2}g-\epsilon}
\right\}+      \nonumber \\
&{}&
\frac{g}{2}\mbox{sin}(\frac{gt}{\sqrt{2}})
\left\{
\frac{\mbox{sin}(\frac{gt}{\sqrt{2}})+
\mbox{sin}((\frac{g}{\sqrt{2}}+\epsilon)t)}{\sqrt{2}g+\epsilon}- 
\frac{\mbox{sin}(\frac{gt}{\sqrt{2}})+
\mbox{sin}((\frac{g}{\sqrt{2}}-\epsilon)t)}{\sqrt{2}g-\epsilon}
\right\},    \nonumber \\
&&{}         \nonumber \\
x_{2}(t)&=&\frac{-i\mbox{sin}(\sqrt{2}gt)}{\sqrt{2}}-
\frac{\sqrt{2}g}{4}
\mbox{sin}(\sqrt{2}gt)\frac{\mbox{sin}(\epsilon t)}{\epsilon}+  
\nonumber \\
&&{}\frac{i\sqrt{2}g}{4}
\left\{
\frac{\mbox{sin}(\epsilon t)+\mbox{sin}(\sqrt{2}g t)}{\sqrt{2}g+\epsilon}+
\frac{\mbox{sin}(\epsilon t)-\mbox{sin}(\sqrt{2}g t)}{\sqrt{2}g-\epsilon}
\right\}-       \nonumber \\  
&&{}\frac{\sqrt{2}g}{8}
\left\{
\frac{\mbox{cos}((\sqrt{2}g+\epsilon)t)-\mbox{cos}(\sqrt{2}g t)}
{2\sqrt{2}g+\epsilon}+
\frac{\mbox{cos}((\sqrt{2}g-\epsilon)t)-\mbox{cos}(\sqrt{2}g t)}
{2\sqrt{2}g-\epsilon}
\right\},    \nonumber \\
&&{}         \nonumber \\
x_{3}(t)&=&\frac{-1+\mbox{cos}(\sqrt{2}gt)}{2}-
\frac{ig}{4}\left(2+\mbox{cos}(\sqrt{2}gt)\right)
\frac{\mbox{sin}(\epsilon t)}{\epsilon}-
\nonumber \\
&&{}\frac{ig}{8}
\left\{
\frac{\mbox{sin}(\sqrt{2}gt)+\mbox{sin}((\sqrt{2}g+\epsilon)t)}
{2\sqrt{2}g+\epsilon}+  
\frac{\mbox{sin}(\sqrt{2}gt)+\mbox{sin}((\sqrt{2}g-\epsilon)t)}
{2\sqrt{2}g-\epsilon}
\right\}+          \nonumber \\
&&{}
\frac{g}{2}\mbox{cos}(\frac{gt}{\sqrt{2}})
\left\{
\frac{-\mbox{cos}(\frac{gt}{\sqrt{2}})+
\mbox{cos}((\frac{g}{\sqrt{2}}+\epsilon)t)}{\sqrt{2}g+\epsilon}-
\frac{-\mbox{cos}(\frac{gt}{\sqrt{2}})+
\mbox{cos}((\frac{g}{\sqrt{2}}-\epsilon)t)}{\sqrt{2}g-\epsilon}
\right\}.         \nonumber 
\end{eqnarray}
We leave the check of this calculation to the readers.

As a result we obtain the approximate solution (up to the next--leading) 
in the case of $n=3$ 
\begin{equation}
\label{eq:approximate-solution}
\Psi(t)=
\left(
  \begin{array}{ccc}
    1&                           &                 \\
     & \mbox{e}^{-i\omega_{1}t}  &                 \\
     &   & \mbox{e}^{-i(\omega_{1}+\omega_{2})t}   \\
  \end{array}
\right)
\left(
  \begin{array}{c}
    x_{1}(t)  \\
    x_{2}(t)  \\
    x_{3}(t)
  \end{array}
\right)=
\left(
  \begin{array}{c}
    \qquad \quad \ \ \ x_{1}(t)                        \\
    \quad \ \ \mbox{e}^{-i\omega_{1}t} x_{2}(t)      \\
    \mbox{e}^{-i(\omega_{1}+\omega_{2})t}x_{3}(t)
  \end{array}
\right)
\end{equation}
with $x_{1}(t), x_{2}(t), x_{3}(t)$ above. This is our main result.

\section{Discussion}

In this paper we developed a general theory of the $n$ level system by 
assuming the rotating wave approximation and gave an analytical solution 
under the ``consistency condition". 
Moreover in the three level system we gave some approximate solution 
(up to the next--leading term), which may be enough because we are 
treating the weak coupling regime ($g$ is small enough compared to the 
energy differences of the atom). 
To continue further calculations (higher order corrections) is not easy 
at the present. 

Here we state our motivation once more. In Quantum Computation with three 
level system we must first of all construct or approximate all elements 
in $U(3)$ by making use of (a kind of) Rabi oscillations $\cdots$ 
a universality of one qudit theory. This is not an easy problem. 
In fact, in almost all papers in qudit theory this has not been mentioned 
as far as we know. 
Our paper is an attempt to this problem and we gave the approximate 
solutions (\ref{eq:approximate-solution}). We need a furthermore study. 

We list the equation in the three level system once more 
\begin{equation}
i\frac{d}{dt}
\left(
  \begin{array}{c}
    \Psi_{1}  \\
    \Psi_{2}  \\
    \Psi_{3}
  \end{array}
\right)
=
\left(
  \begin{array}{ccc}
    0 & g\mbox{e}^{i\omega_{1}t} & g\mbox{e}^{i\omega_{02}t}  \\
    g\mbox{e}^{-i\omega_{1}t} & \Delta_{1} & g\mbox{e}^{i\omega_{2}t}  \\
    g\mbox{e}^{-i\omega_{02}t} & g\mbox{e}^{-i\omega_{2}t} & \Delta_{2} 
  \end{array}
\right)
\left(
  \begin{array}{c}
    \Psi_{1}  \\
    \Psi_{2}  \\
    \Psi_{3}
  \end{array}
\right). 
\end{equation}
and present 

\par \vspace{3mm} \noindent 
{\bf Problem}\quad Show a good method to solve (or to solve approximately) 
the equation above. 

This problem seems to be very hard.  

\par \noindent 
Moreover, study the general equation in the three level system 
(without the rotating wave approximation) 
\begin{equation}
i\frac{d}{dt}
\left(
  \begin{array}{c}
    \Psi_{1}  \\
    \Psi_{2}  \\
    \Psi_{3}
  \end{array}
\right)
=
\left(
  \begin{array}{ccc}
   0 & g\mbox{cos}(i\omega_{1}t) & g\mbox{cos}(i\omega_{02}t)  \\
   g\mbox{cos}(-i\omega_{1}t) & \Delta_{1} & g\mbox{cos}(i\omega_{2}t)  \\
   g\mbox{cos}(-i\omega_{02}t) & g\mbox{cos}(-i\omega_{2}t) & \Delta_{2} 
  \end{array}
\right)
\left(
  \begin{array}{c}
    \Psi_{1}  \\
    \Psi_{2}  \\
    \Psi_{3}
  \end{array}
\right). 
\end{equation}
This is our next target !

\vspace{5mm}
\noindent{\em Acknowledgment.}\\
K. Fujii wishes to thank Akira Asada, Marco Frasca, Kunio Funahashi and 
Shin'ichi Nojiri for their helpful comments and suggestions.

\newpage


\end{document}